\documentclass[twocolumn,english,floatfix,preprintnumbers,showpacs,amsfonts,amssymb,superscriptaddress]{revtex4}
\usepackage{ae,aecompl}
\usepackage[T1]{fontenc}
\usepackage[latin1]{inputenc}
\usepackage{amsmath}
\usepackage{amssymb}
\usepackage{graphicx}

\makeatletter
\@ifundefined{textcolor}{}
{%
 \definecolor{BLACK}{gray}{0}
 \definecolor{WHITE}{gray}{1}
 \definecolor{RED}{rgb}{1,0,0}
 \definecolor{GREEN}{rgb}{0,1,0}
 \definecolor{BLUE}{rgb}{0,0,1}
 \definecolor{CYAN}{cmyk}{1,0,0,0}
 \definecolor{MAGENTA}{cmyk}{0,1,0,0}
 \definecolor{YELLOW}{cmyk}{0,0,1,0}
}

\usepackage{amscd}
\usepackage{bm}
\usepackage{graphics,psfrag}
\usepackage{graphicx,psfrag}
\usepackage{lipsum}

\makeatother

\usepackage{babel}
\begin{document}

\title{Universal Finite-Size Corrections of the Entanglement Entropy of
Quantum Ladders and the Entropic Area Law }

\author{J.~C.~Xavier }

\affiliation{Universidade Federal de Uberlândia, Instituto de F\'{\i}sica, Caixa
Postal 593, 38400-902 Uberlândia, MG, Brazil }

\author{F.~B.~Ramos }

\affiliation{Universidade Federal de Uberlândia, Instituto de F\'{\i}sica, Caixa
Postal 593, 38400-902 Uberlândia, MG, Brazil }

\date{\today{}}
\begin{abstract}
We investigate the finite-size corrections of the entanglement entropy
of critical ladders and propose a conjecture for its scaling behavior.
The conjecture is verified for free fermions, Heisenberg and quantum
Ising ladders. Our results support that the prefactor of the logarithmic
correction of the entanglement entropy of critical ladder
models is universal and it is associated with the central charge of
the one-dimensional version of the models and with the number of branches
associated with gapless excitations. Our results suggest that it is possible
to infer whether there is a violation of the entropic area law in
two-dimensional critical systems by analyzing the scaling behavior
of the entanglement entropy of ladder systems, which are easier to
deal.
\end{abstract}


\maketitle
\emph{Introduction.} Entanglement is a very peculiar property of composite
systems which has intrigued the physicists since the beginning of
quantum mechanics. The entanglement is a fundamental ingredient to
teleport quantum states and it is also an important key in quantum
computation and quantum information \citep{Horodeckirev}. Among the
various quantifiers of entanglement, the entanglement entropy (EE)
is one of the most used since it is sensitive to the long-distance
quantum correlations of critical systems. 

In the last years, physicists working in distinct areas (such as quantum
information, quantum field theory and condensed matter) have made
a great effort to understand the scaling behavior of the EE of bipartite
systems. In particular, the violation of the entropic area law has
been a highly debated issue in recent years \citep{revfazio,PhysRevLett.71.666,PhysRevLett.94.060503,PhysRevLett.96.010404,PhysRevLett.96.100503,PhysRevB.74.073103,PhysRevA.74.022329,PhysRevLett.105.050502,farkas,prl112-160403,EPL97-20009,RMP82-277,PhysRevLett.100.215701,cvidal,entrosoc2}.
The EE of two composite subsystems ${\cal {A}}$ and ${\cal {B}}$
is defined as the von Neumann entropy $S_{{\cal {A}}}=-\mbox{Tr}\rho_{{\cal {A}}}\ln\rho_{{\cal {A}}}$,
associated to the reduced density matrix $\rho_{{\cal {A}}}=\mbox{Tr}_{{\cal {B}}}\rho$.
Since $S_{{\cal {A}}}=S_{{\cal {B}}}$, the information is shared
only among the degrees of freedom localized around the surface (``area'')
separating both systems, due to this fact it is expected that the
EE of cube ${\cal {A}}$ with side ${\cal {N}}$  behaves as 
$S_{{\cal {A}}}\sim{\cal {N}}^{d-1}$,
where $d$ is the dimension and ${\cal {N}}^{d-1}$ 
is the boundary "area" separating the regions ${\cal {A}}$ and ${\cal {B}}$. 
Indeed, this scaling behavior is expected
for gapped systems \citep{PRL100-070502} and was also observed for
some critical systems (see Ref. \onlinecite{RMP82-277} and references
therein). On the other hand, some models such as the one-dimensional
critical systems \citep{cardyentan}, the free fermions systems with
a finite Fermi surface in any dimension \citep{PhysRevB.74.073103,PhysRevA.74.022329},
the two-dimensional (2D) Heisenberg model \citep{MelkoPRB84-165134,RachelPRB-83-224410}
and the 2D conformal critical systems \citep{FradkinPRL97-050404,PasquierPRB80-184421}
present beyond the ${\cal {N}}^{d-1}$  correction a logarithmic term.

\begin{figure}[!]
\begin{centering}
\includegraphics[clip,width=4cm]{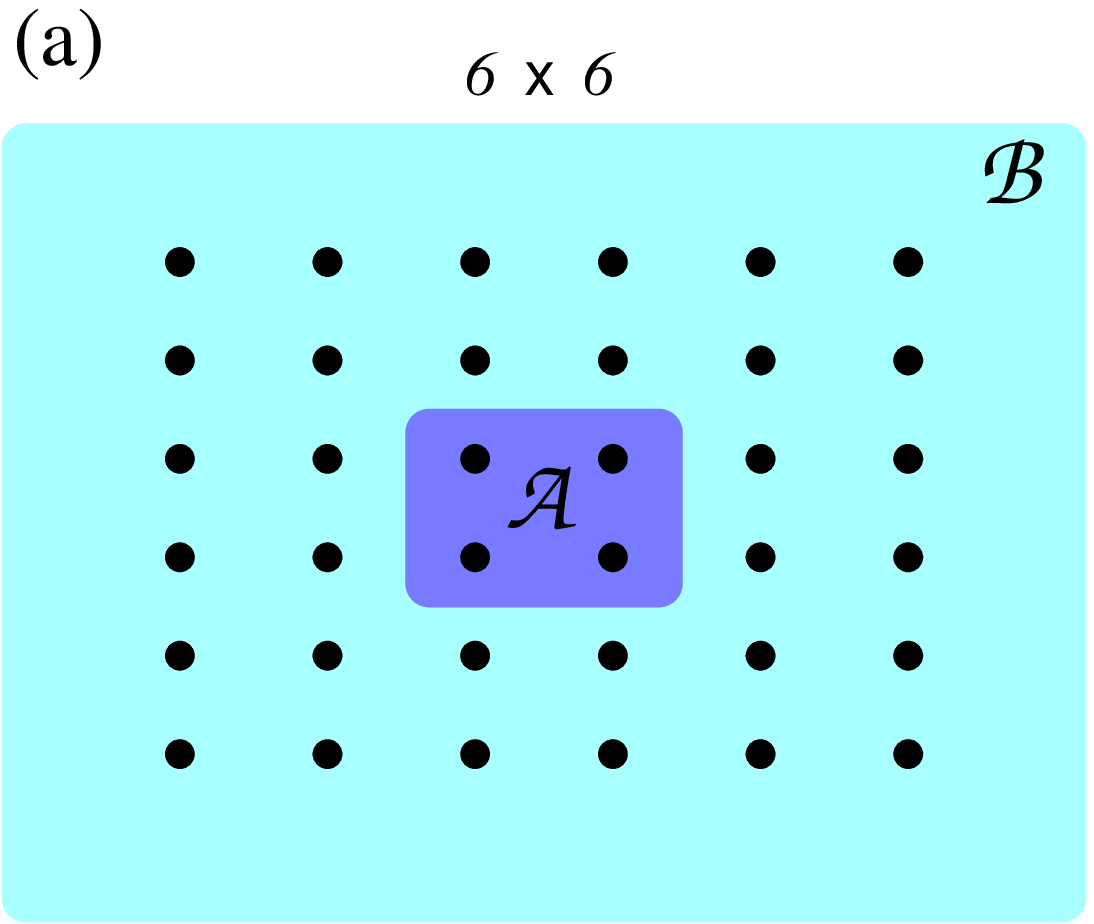} \includegraphics[clip,width=4cm]{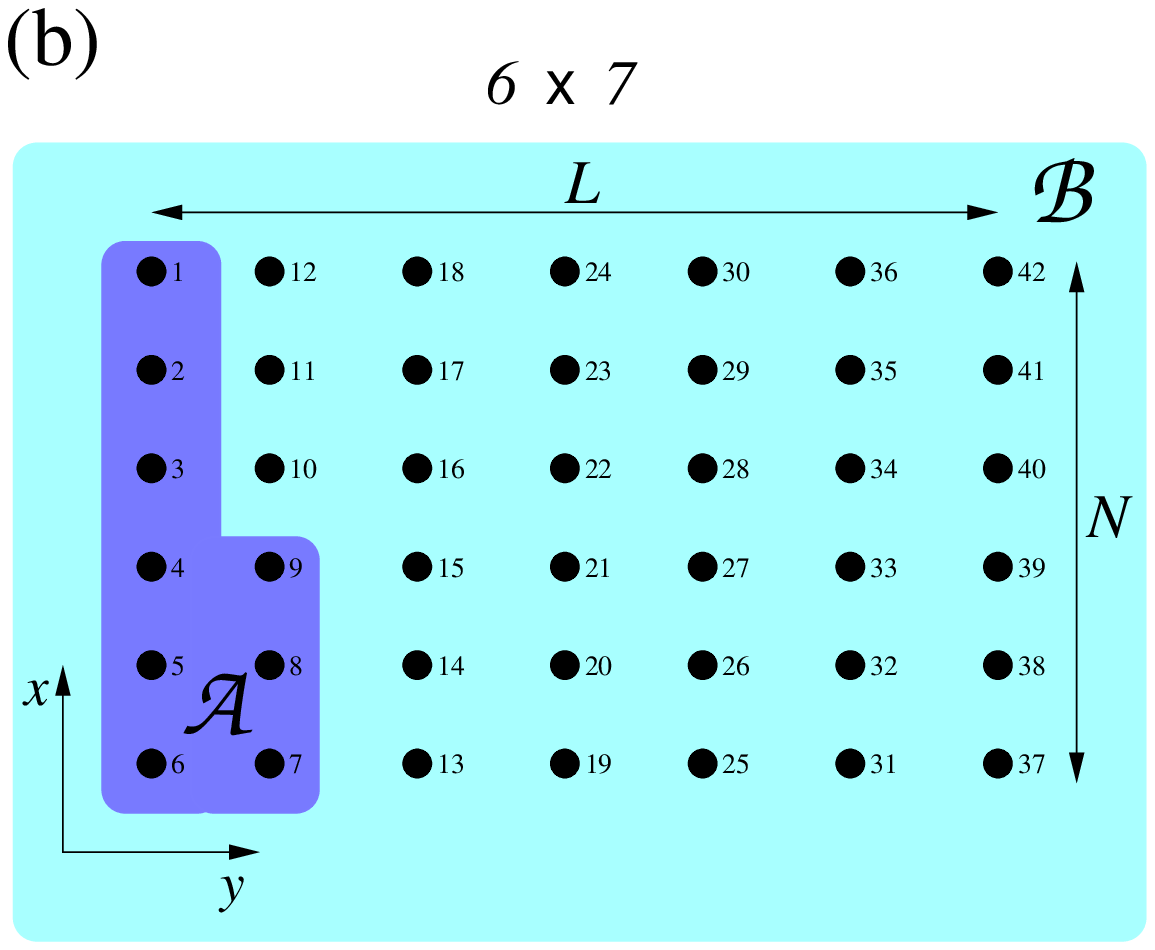}
\par\end{centering}

\caption{\label{fig1}(Color online) Illustration of  six-leg ladders divided
into two entangled blocks. In (a) the subsystem ${\cal {A}}$ is immersed
in the middle of the system while in (b) the subsystem ${\cal {A}}$
is in corner of the ladder. We also present the labels of the sites. }
\end{figure}

It is well known that the prefactor of the logarithmic correction
of critical one-dimensional systems of size $L$ is universal and
it is associated with the central charge $c$ by the following equation
\citep{cardyentan}

\begin{equation}
S(L,\ell)=\frac{c}{3\eta}\ln\left[\frac{\eta L}{\pi}\sin\left(\frac{\pi\ell}{L}\right)\right]+a,\label{eq:1dcentralcharge}
\end{equation}
where $\ell$ is the size of the subsystem ${\cal {A}}$, $a$ is
a non-universal constant and $\eta=1$(2) for the systems under periodic
(open/fixed) boundary conditions. Note that other subleading corrections
exist and are related with the scaling dimensions \citep{entropyosc}.

For any dimension $d$, it is expected the following general behavior
for the EE of a cube ${\cal {A}}$ with side ${\cal {N}}$ (see Fig. \ref{fig1})

\begin{equation}
S(\ell)=A{\cal {N}}^{d-1}+C({\cal {N}})\ln\left({\cal {N}}\right)+B.\label{eq:2DEE}
\end{equation}
In this work, we determine numerically $C({\cal {N}})$ for some \emph{quantum
ladders} and found that it is universal. The $N$-leg ladders are
characterized by $N$ parallel chains of size $L$ coupled one to each others \citep{dagottosc1}.
We denote the size of the ladders by $N\times L$.
The $N$-leg ladders are easier to deal than the two-dimensional systems
and can be used as a simple route to study the EE of the two-dimensional
systems. Here, we consider ladders composed of the following critical
chains: free fermions chains, Heisenberg chains and the quantum Ising
chains. 

Although most of the works done in the literature consider the subsystem
${\cal {A}}$ immersed in a ``reservoir'', as illustrated in the
Fig. \ref{fig1}(a), for ladder systems is convenient to consider
the subsystem ${\cal {A}}$ in the corner of the ladders {[}see Fig.
\ref{fig1}(b){]}. Our main aim is to present a conjecture to the
scaling behavior of the EE of critical ladders. Surprisingly, we verify
that the finite-size corrections of the EE of quantum ladders are
very similar to those of critical chains {[}Eq. (\ref{eq:1dcentralcharge}){]}.
Consider a ladder system composed of $N$ quantum chains of size $L$,
and let $\ell$ be the number of sites of the block ${\cal {A}}$
labeled as Fig. \ref{fig1}(b). We propose that the scaling behavior
of the EE of critical ladders is given by

\[
S(\ell)=AN+\frac{c}{3\eta_{x}}N_{gl}\ln\left[\frac{\sin\left(\frac{\pi\ell}{NL}\right)}{\sin\left(\frac{\pi}{L}\right)}\right]+B
\]
\begin{equation}
+\sum_{j=1}^{[\frac{N}{2}]}a_{j}\cos\left(2\pi\ell j/N\right)\,,\label{eq:3conjecture}
\end{equation}
where $c$ is the central charge (of the quantum chain used to build
the ladders), $N_{gl}$ is the number of dispersion branches associated
with the gapless excitations for a given energy, 
$\eta_{x}$=1
(2) for ladders under periodic (open/fixed) boundary in the $x$ direction
and $A$, $B$ and $a_{j}$ are non-universal constants. The last
term in the above equation is an ansatz that we use which has been
shown to be efficient for describing the oscillations of the EE. 
The importance of the number of gapless modes in the
the EE have been discussed in spin systems \cite{gaplesspin} and
boson systems. \cite{gaplessboson}
The above conjecture indicates that the prefactor of the logarithmic correction
of the EE of \emph{critical ladders} is universal
and it is related with the universality class of critical behavior
of the chains that are used to build the quantum ladders. Note that
for gapped systems $N_{gl}=0$ and the Eq. (\ref{eq:3conjecture})
suggests us that the entropic area law holds in this case, as expected.
Below, we present results for critical ladders that support our conjecture. 

\begin{figure}[!]
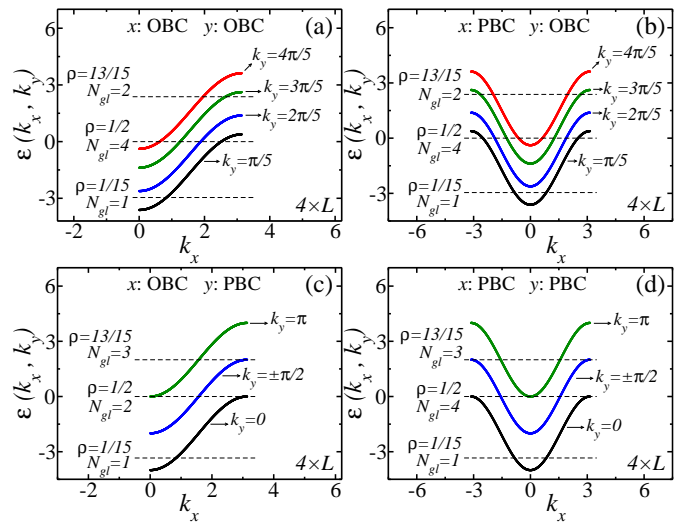

\begin{centering}
\includegraphics[clip,width=4.4cm]{fig2a.eps}\includegraphics[clip,width=4.4cm]{fig2b.eps}
\par\end{centering}

\begin{centering}
\includegraphics[clip,width=4.4cm]{fig2c.eps}\includegraphics[clip,width=4.4cm]{fig2d.eps}
\par\end{centering}

\caption{\label{fig2}(Color online) The band dispersions of the four-leg free
fermions ladders for different boundary conditions. The horizontal dashed
lines indicate the positions of the Fermi levels for three values of
densities $\rho$. We also indicate the values of $N_{gl}$ associated
with each density. Note that some branches are degenerate.}
\end{figure}

\begin{figure}[!]
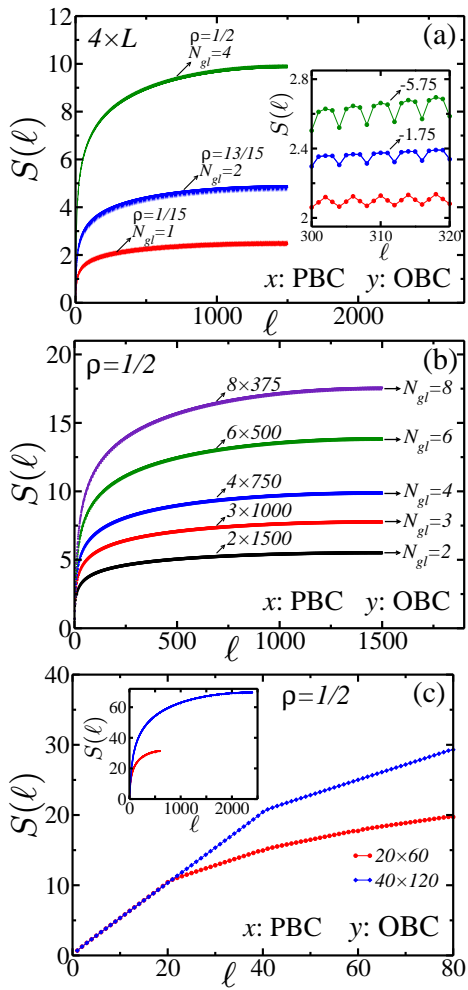

\begin{centering}
\psfrag{axis1}{\scalebox{1.5}{$\ell$}}
\psfrag{axis2}{\scalebox{1.5}{$S(\ell)$}}
\psfrag{axis3}{\scalebox{1.0}{$\ell$}}
\psfrag{axis4}{\scalebox{1.0}{$S(\ell)$}}
\includegraphics[clip,width=6cm]{fig3a.eps}
\par\end{centering}

\begin{centering}
\psfrag{axis1}{\scalebox{1.5}{$\ell$}}
\psfrag{axis2}{\scalebox{1.5}{$S(\ell)$}}
\includegraphics[clip,width=6cm]{fig3b.eps}
\par\end{centering}

\begin{centering}
\psfrag{axis1}{\scalebox{1.5}{$\ell$}}
\psfrag{axis2}{\scalebox{1.5}{$S(\ell)$}}
\psfrag{axis3}{\scalebox{1.0}{$\ell$}}
\psfrag{axis4}{\scalebox{1.0}{$S(\ell)$}}
\includegraphics[clip,width=6cm]{fig3c.eps}
\par\end{centering}

\caption{\label{fig3}(Color online) $S(\ell)$ vs. $\ell$ for the free fermions
ladders. (a) Results for a cluster $4\times750$ and three values
of $\rho$. Inset: $S(\ell)$ for few sites. In order to show all
data in the figure we added some constants in the values of $S$.
(b) Data of the EE for several ladders at half-filling. From these
fits we get $A=0.56$, $B=0.37$. The non-universal constants $a_{j}$
are small and varying from $-0.04$ to $-0.01$. (c) Results for the
twenty- and forty-leg ladders at half-filling. In (a) and (b) the
symbols are the data obtained by the correlation matrix method (see text)
and the solid lines connect the fitted points by using our conjecture
{[}Eq. \ref{eq:3conjecture}{]}. }
\end{figure}

\emph{Free Fermions Ladders.} Let us first consider a free-fermions
ladders whose Hamiltonian is given by 
\begin{equation}
H=\sum_{k_{x},k_{y}}\mathcal{E}(k_{x},k_{y})c_{k_{x},k_{y}}^{\dagger}c_{k_{x},k_{y}}\:,\label{eq:hamiltonian}
\end{equation}

\noindent where the dispersion is $\mathcal{E}(k_{x},k_{y})=-2\left[\cos(k_{x})+\cos(k_{y})\right]$
and  sum is 
taken for all  wave numbers in the Brillouin zone.
The momenta are given by $k_{x}=j_{x}\frac{2\pi}{L}$
{[}$j_{x}\frac{\pi}{L+1}$ {]} and $k_{y}=j_{y}\frac{2\pi}{N}$ {[}$j_{y}\frac{\pi}{N+1}$
{]} for periodic {[}open{]} boundary condition in $x$ and $y$ directions,
respectively. The variables $j_{x}$ and $j_{y}$ are integers and
its values depend on the boundary conditions.

\noindent In the case of free fermions systems it is possible to determine
the EE for very large systems by using the correlation matrix method
\citep{0305-4470-36-14-101}. Note that in principle it is possible
to use the Widom conjecture \citep{PhysRevLett.96.100503,widom} to
determine the prefactor that appears in the logarithmic correction
(see for example Ref. \onlinecite{EPL97-20009}). However, we observe
that this prefactor is easier to understand in terms of the number
of gapless modes $N_{gl}$ that cross the Fermi level. For the sake
of clarification, we display in Fig. \ref{fig2} the band dispersions
for the four-leg ladder as well as the values of $N_{gl}$ for some
densities \textbf{$\rho$}. For the half-filling case with periodic
boundary condition (PBC) in the $x$ direction and open boundary condition
(OBC) in $y$ direction, the number of gapless modes that cross
the Fermi level is equal to the number of legs, i. e., $N_{gl}=N$
(for the other boundary conditions $N_{gl}\approx N$ for large values
of $N$). So, based in our conjecture we expect that the EE for large
values of $N$ and $L$ behaves as $S(\ell=NL/2)=AN+\frac{1}{6}N\ln(\frac{L}{\pi})+B$,
which suggest that the entropic area law is broken for the half-filling
case. Indeed, this was observed in free fermions systems in two dimensions 
\citep{PhysRevLett.96.100503,PhysRevLett.96.010404,PhysRevB.74.073103,entroschollwock,prl112-160403}. 

\noindent In Fig \ref{fig3}(a), we present $S(\ell)$ as function
of $\ell$ for a cluster of size $4\times750$ with PBC {[}OBC{]}
in the $x$ {[}$y${]} direction and three values of densities. As
we observe, the data obtained by the correlation matrix method agree perfectly
with the conjecture proposed {[}Eq. (\ref{eq:3conjecture}){]}. In the
fitting procedure, we
used $c=1$ (which corresponds to the central charge of the one-dimensional
chain) and the values of $N_{gl}$ used were obtained counting the
number of gapless modes that cross the Fermi level, as illustrated
in Fig. \ref{fig2}. Similar agreements are found for several other
ladders, as shown in Fig. \ref{fig3}(b).

In order to understand the contribution of the first term of Eq. (\ref{eq:3conjecture}),
we present in Fig. \ref{fig3}(c) the EE for the $20\times60$ and
the $40\times120$ clusters with PBC {[}OBC{]} in the $x$ {[}$y${]}
direction at half-filling. As we can note, $S(\ell)$ grows lineary
for $\ell\le N$ and the logarithmic scaling is present only for $\ell\ge N$
{[}see inset of Fig. \ref{fig3}(c){]}. If we impose an ansatz for
$S(\ell)$ similar to the Eq. (\ref{eq:1dcentralcharge}) and use the
fact that $S(\ell)$ is continuous at $\ell=N$ 
(i. e., $AN+B=\frac{c}{3\eta_{x}}N_{gl}\ln\left[\frac{NL}{\pi}\sin\left(\frac{\pi}{L}\right)  \right]+a$)
we realize that the EE must behave as Eq. (\ref{eq:3conjecture}). This is very interesting,
since in principle we can obtain the prefactor $A$ by studying the
behavior of $S(\ell)$ for $\ell<N$, which is easier to obtain. 

\begin{figure}[!]
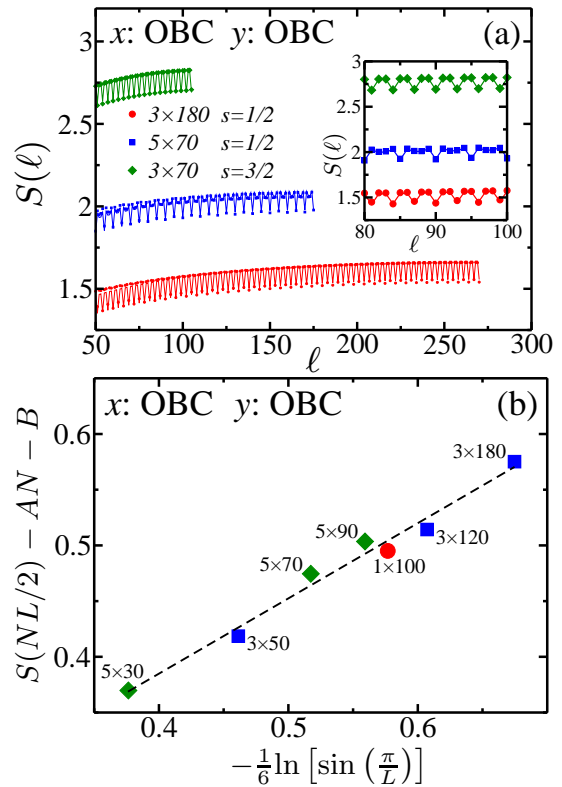

\begin{centering}
\psfrag{axis1}{\scalebox{1.5}{$\ell$}}
\psfrag{axis2}{\scalebox{1.5}{$S(\ell)$}}
\psfrag{axis3}{\scalebox{1.0}{$\ell$}}
\psfrag{axis4}{\scalebox{1.0}{$S(\ell)$}}
\includegraphics[clip,width=7cm]{fig4a.eps}
\end{centering}
\begin{centering}
\psfrag{axis1}{\scalebox{1.27}{$-\frac{1}{6}\mathrm{ln}\left[\mathrm{sin}\left(\frac{\pi}{L}\right)\right]$}}
\psfrag{axis2}{\scalebox{1.27}{$S(NL/2)-AN-B$}}
\includegraphics[clip,width=7cm]{fig4b}
\end{centering}

\caption{\label{fig4}(Color online) (a) $S(\ell)$ for the Heisenberg ladders
with spins $s=1/2$ and $s=3/2$. The symbols are the data obtained
by DMRG and the solid lines connect the fitted points by using our
conjecture {[}Eq. \ref{eq:3conjecture}{]} with $c=1$ and $N_{gl}=1$.
From these fits we get $A=0.27$ and  $B=0.16$ for $s=1/2$.  
Inset shows $S(\ell)$ for few sites. (b) 
$S(\ell=NL/2)-AN-B$ vs. $-1/6\ln\left[ \sin\left(\frac{\pi}{L}\right) \right]$
for several cluster sizes with $s=1/2$. }
\end{figure}

\emph{Heisenberg Ladders.} Now, let us consider the $N$-leg spin-$s$
Heisenberg ladders whose hamiltonian is given by 

\[
H=J\sum_{i=1}^{N}\sum_{j=1}^{L-1}\mathbf{S}_{i,j}\cdot\mathbf{S}_{i,j+1}+J\sum_{i=1}^{N-1}\sum_{j=1}^{L}\mathbf{S}_{i,j}\cdot\mathbf{S}_{i+1,j}\:,
\]
where $\mathbf{S}_{i,j}$ is the spin-$s$ operator at the $i$-th
leg and $j$-th rung. We have set $J=1$ to fix the energy scale.
It is well known that the $N$-leg spin-$s$ Heisenberg ladders is
gapless (gapped) if $sN$ is semi-integer (integer) \citep{dagottosc1,haldaneconj1},
see also the Ref. \onlinecite{flaviaXavierNlegspin} and references therein.
Here, we focus in the case of critical ladders, i. e. $sN$ is semi-integer.
For the Heisenberg ladders case, we obtained numerically the EE by using the density-matrix
renormalization group (DMRG) \citep{white}. For simplicity we consider
only OBC in both directions. The spin-$s$ Heisenberg
chains with semi-integer spins have central charge $c=1$ \citep{spinsaffleck}.
Besides, based in the spin wave approximation it is expected that
the dispersion of the 2D Heisenberg model has one Goldstone mode $E(k)\sim\sqrt{k_{x}^{2}+k_{y}^{2}}$.
Since the number of legs $N$ is finite, the values of $k_{y}$ are
discrete. Due to this fact, in analogous to the free fermions case,
there is just one dispersion branch ($E(k_{x},0)\sim|k_{x}|$) associated
with gapless excitations that crosses the energy of the ground state,
i.e. $N_{gl}=1$. In Fig. \ref{fig4}(a), we display the $S(\ell)$
as function of $\ell$ for the Heisenberg ladders with spins $s=1/2$ and
$s=3/2$. Similar to the free fermions case, the Eq. (\ref{eq:3conjecture})
reproduces quite well the scaling behavior of $S(\ell)$ if we use
$c=1$ and $N_{gl}=1$. Note that in this case, our results suggest that a violation of the entropic area law
is not expected in the two-dimensional systems.
The EE for large values of $N$
and $L$ should behave as $S(\ell=NL/2)=AN+\frac{1}{6}\ln(\frac{L}{\pi})+B$.
In order to verify this, we present in  Fig. \ref{fig4}(b)
 $S(\ell=NL/2)-AN-B$ as function of $-1/6\ln\left[ \sin\left(\frac{\pi}{L}\right) \right] $. 
As we see, the data strongly indicate that the prefator of the logarithmic
term is 1/6 for the Heisenberg ladders when the subsystem is in the conner. 
Note that this result is intriguing, at least for the point of view of
$N$ uncoupled chains under OBC, which could suggest that the prefactor is $N/6$.
Note that Monte Carlo simulations \citep{MelkoPRB84-165134} as well
as the DMRG results \citep{RachelPRB-83-224410} show a similar behavior for 
the scaling of the EE for other aspect ratio.

\begin{figure}[!]
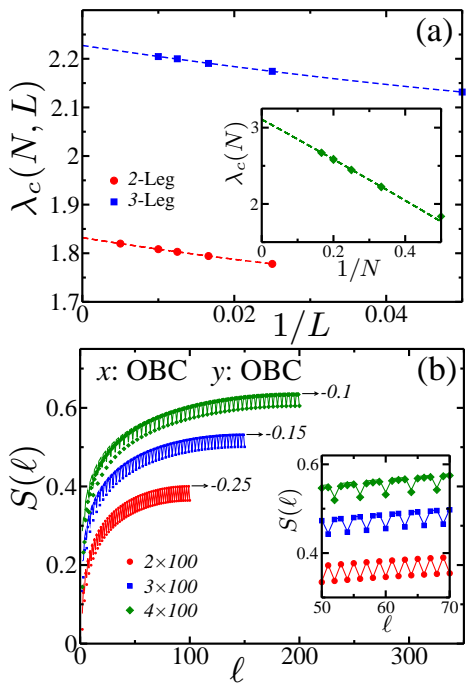

\begin{centering}
\psfrag{axis1}{\scalebox{1.45}{$1/L$}}
\psfrag{axis2}{\scalebox{1.5}{$\lambda_c(N,L)$}}
\psfrag{axis3}{\scalebox{1.0}{$1/N$}}
\psfrag{axis4}{\scalebox{1.0}{$\lambda_c(N)$}}
\includegraphics[clip,width=6cm]{fig5a.eps}
\par\end{centering}

\vspace{1mm}
\begin{centering}
\psfrag{axis1}{\scalebox{1.5}{$\ell$}}
\psfrag{axis2}{\scalebox{1.5}{$S(\ell)$}}
\psfrag{axis3}{\scalebox{1.0}{$\ell$}}
\psfrag{axis4}{\scalebox{1.0}{$S(\ell)$}}
\includegraphics[clip,width=6cm]{fig5b.eps}
\par\end{centering}

\caption{\label{fig5}(Color online) (a) Finite-size estimates of the critical
point, $\lambda_{c}(N,L)$, as function of $1/L$ for the two- and
three-leg Ising ladders. Inset: $\lambda_{c}^{N}$ vs. $1/N$. (b)
$S(\ell)$ vs. $\ell$ for three values of $N$ at the critical points.
The symbols are the DMRG results and the solid lines connect the fitted
points by using our conjecture {[}Eq. \ref{eq:3conjecture}{]} with
$c=1/2$ and $N_{gl}=1$.  In order to show all
data in the figure we added some constants in the values of $S$.
Inset shows $S(\ell)$ for few sites. }
\end{figure}

\emph{Quantum Ising Ladders.} Finally, let us consider the $N$-leg
quantum Ising ladders whose hamiltonian is given by

\[
H=\sum_{i=1}^{N}\sum_{j=1}^{L-1}\sigma_{i,j}^{x}\sigma_{i,j+1}^{x}+\sum_{i=1}^{N-1}\sum_{j=1}^{L}\sigma_{i,j}^{x}\sigma_{i+1,j}^{x}+\lambda\sum_{i=1}^{N}\sum_{j=1}^{L}\sigma_{i,j}^{z}\:,
\]
where $\sigma_{i,j}^{x,y,z}$ are Pauli matrices at the $i$-th leg
and $j$-th rung. The one-dimensional case, i. e. $N=1$, has a critical
point at $\lambda_{c}=1$ and its critical behavior is described by a
conformal field theory with central charge $c=1/2$.
In order to test the validity of Eq. (\ref{eq:3conjecture}) for the
Ising ladders, we have first to determine the critical values of $\lambda_{c}^{N}$
for each value of $N$. First, we get the finite-size estimates of
$\lambda_{c}(N,L)$ using the EE as reported in Ref. \onlinecite{xavieralcarazQCP}.
Then, we assume that $\lambda_{c}(N,L)$ behaves as $\lambda_{c}(N,L)$=
$\lambda_{c}(N)+a/L+b/L^{2}$, and finally we fit the data to obtain
$\lambda_{c}(N)$. As illustration, we present in Fig. \ref{fig5}(a)
$\lambda_{c}(N,L$) as function of $1/L$ for the two and three-leg
Ising ladders. By fitting our data we obtained $\lambda_{c}(N)$ = 1.838,
2.219, 2.443, 2.578, and 2.670 for $N=2,3,4,5$ and 6, respectively.
It is interesting to note that if we extrapolate these estimates to obtain
$\lambda_{c}(\infty)$, as report in the inset of Fig. \ref{fig5}(a),
we obtain $\lambda_{c}^{2D}=\lambda_{c}(\infty)=3.1$, which is close
to the estimates of the critical point of the two-dimensional quantum
Ising model obtained by Monte Carlo \citep{2DisingPRE66} $(\lambda_{c}^{2D}=3.044)$
and by the multiscale entanglement renormalization ansatz \citep{VidalPRL102-180406}
$(\lambda_{c}^{2D}=3.07)$. The small discrepancy between our estimate
and the last ones  is very probable associated with the small lattice sizes
considered to extrapolate our data. 

As in the Heisenberg model, it is expected that $N_{gl}=1$ for the
critical Ising ladders, and we do not anticipate a violation of the
entropic area law for the two-dimensional quantum Ising model. The
EE should behaves, at the critical point, as $S(\ell=NL/2)=AN+\frac{1}{12}\ln(\frac{L}{\pi}) +B$,
for OBC in both directions. In Fig. \ref{fig5}(b), we present the
EE of the Ising ladders at the critical points acquired by DMRG for
$N=2,3$ and $N=4$. As we can note in this figure, the conjecture
proposed {[}Eq. (\ref{eq:3conjecture}){]} also reproduces quite well
the scaling behavior of the EE of the critical Ising ladders.

\emph{Conclusions.} We present an ansatz {[}Eq. (\ref{eq:3conjecture}){]}
for the finite-size corrections of the entanglement entropy of critical
ladders. We verify that the ansatz is able to reproduce quite well
the scaling behavior of the entanglement entropy of some critical
ladders, namely: free fermions ladders, Heisenberg ladders and Ising
ladders. Preliminary results of the quantum $q=3$ Potts ladders (not
shown) also corroborate with the scaling behavior of the entanglement
entropy proposed. All those  results support that the prefactor
of the logarithmic correction of the critical ladders
is universal and it is related with central charge
of the one-dimensional version of the model as well as the number
of branches associated with gapless excitations. Note that Eq. 3 is valid for  
$L>>N$ and only when the subsystem ${\cal {A}}$ is consider in the  corner of the ladder. 
A puzzle still unsolved, is find the exact value of the prefactor of the  
logarithmic term, when the subsystem  ${\cal {A}}$ is immersed in the middle of the ladders.
\begin{acknowledgments}
The authors thank B. Bauer to point out the Ref. \onlinecite{gaplessboson}.
This research was supported by the Brazilian agencies FAPEMIG and CNPq. 
\end{acknowledgments}

\end{document}